**Learning in a Landscape: Simulation-building as Reflexive Intervention**

*Version 2010*

**Anne Beaulieu, Matt Ratto, and Andrea Scharnhorst**

**Abstract**

This article makes a dual contribution to scholarship in science and technology studies (STS) on simulation-building. It both documents a specific simulation-building project, and demonstrates a concrete contribution to interdisciplinary work of STS insights. The article analyses the struggles that arise in the course of determining what counts as theory, as model and even as a simulation. Such debates are especially decisive when working across disciplinary boundaries, and their resolution is an important part of the work involved in building simulations. In particular, we show how ontological arguments about the value of simulations tend to determine the direction of simulation-building. This dynamic makes it difficult to maintain an interest in the heterogeneity of simulations and a view of simulations as unfolding scientific objects.

As an outcome of our analysis of the process and reflections about interdisciplinary work around simulations, we propose a chart, as a tool to facilitate discussions about simulations. This chart can be a means to create common ground among actors in a simulation-building project, and a support for discussions that address other features of simulations besides their ontological status. Rather than foregrounding the chart's classificatory potential, we stress its (past and potential) role in discussing and reflecting on simulation-building as interdisciplinary endeavor. This chart is a concrete instance of the kinds of contributions that STS can make to better, more reflexive practice of simulation-building.

## Ambitions and Struggles in Building Simulations

Could a simulation be based on the rich insights that come out of ethnographic fieldwork? What if some aspects of fieldwork could be represented in a radically different way? What if people's ability to learn and act in new ways could be modeled? These three very different questions came together six years ago, and fuelled a research proposal. The contrasts between these starting points were significant, and they were clear to us from the beginning. We indeed expected challenges and negotiations during the project. What we didn't expect though, was that in the course of building a simulation, we would end up with a whole hard drive full of them! Nor did we envisage that we would so strongly disagree about what counted as proper simulations. One participant's simulation was another participant's game. One researcher's sound underlying model was another's unacceptable reduction. We therefore ended up with a wealth of simulations in different stages of development, which were furthermore very unevenly appreciated within the team. Most importantly, however, we also developed ways to exchange about and understand these differences. We use the notion of "simulation-building" in this paper to point to the *process* of creating simulation tools, some of which were made publicly available. This process entails discussion and negotiation about conceptualization, theoretical model-building, the development of mathematical models, numerical methods to explore them, and, eventually building interfaces that allow users to engage with simulations.

This article explores these expected and unexpected tensions in simulation-building, in order to make a dual contribution. First, what follows is a reflection on the process of simulation-building, and contributes to STS scholarship on simulation in scientific research. Simulation, as a practice, is receiving more attention of late, but case studies of this kind, documenting a specific simulation building project, remain scarce. Second, this account is used to articulate a concrete contribution that STS scholarship can make to simulation-building. As an outcome of our analysis of and reflections about the process, we propose a chart, as a tool to facilitate discussions about simulations. This tool, we will show, can be a means to create common ground among actors in a simulation-building project, and constitutes an effective deployment of STS insights. Rather than foregrounding its classificatory potential, we want to highlight its function as a support for discussing and reflecting on simulation-building as interdisciplinary endeavors.

Two accounts are therefore entwined in this article. We present the struggles that arose in the course of the project, and make explicit the tools, vocabularies and practices we developed to work through them. We want to show that what counts as theory, as model, and as simulation are part of the important work involved in building simulations. We further demonstrate how a chart can help make explicit 'what counts' for both scholars and builders of simulation. By conjoining these two perspectives in this article, we wish to link simulation-building and analysis as a kind of STS-on-the-ground, a position that we want to clarify and distinguish from a positivistic approach that dominates much simulation-building efforts. But before detailing the simulation-building case and explaining how the chart came about, we describe some of the background of the project and of this article.

*The Competence Project*

The simulation-building project aimed to explore a model of learning, the 'competence model'. This 'competence model' has been the object of publications in education and management theory, as part of a larger debate around life-long learning and the role of professional education in Germany (Erpenbeck & Heyse, 1999). Here “model” refers to a theoretical framework, rooted in philosophy and psychology, in which four types of competences have been proposed as most important for human behavior in processes of problem solving and learning: personality, social-communicative competences, knowledge (factual), and activity (as the impetus to act). Empirically operationalised in the form of a questionnaire, the approach had been tested in corporate settings, as a tool for self- and external evaluation. Erpenbeck, the main proponent of this body of work was aiming to link this conceptual framework to mathematical models of complex systems. This may not seem to be an obvious link. However, problem solving has been a topic of research for areas like operations research, engineering and design—all areas where mathematical models are dominant. In this approach, problem solving conceptualized as finding the ‘optimum’ point. Extending this concept with a spatial metaphor, solving a problem would be finding the highest point in a space, and learning could be described as seeking such a ‘high’ location. Mathematical models can be used to describe such optimum points, and to describe strategies to reach these high points. Furthermore, attributes that affect learning can also be expressed as part of a mathematical model—what we will refer to as ‘knowledge spaces’ later on. When applying such an approach to a theory of learning, it is possible to use a mathematical model to describe a situation (what can be learned) and an actor’s behaviour (how one learns). These are the elements needed to produce a spatial rendering, in the shape of a simulation. When building a simulation within such a space, the behaviour of actors can further be modeled on ‘searching’ strategies, which are rendered as the exploration of a space. These strategies can be more or less complex, depending on the attributes of agents (for example, all-seeing or having limited vision) and the complexity of the space (one optimum point, or several ‘hills’).

Often, search strategies are described using elements from evolutionary theories. Computing is used to handle the underlying mathematical models, and to visualise the outcomes of computation into a spatial simulation on a screen.

To return to the goals of our project, Erpenbeck hoped that social-psychological models of competence could also be enhanced by linking and translating them into mathematical models, like those used to conceptualize learning as a process of problem solving and optimizing. He further hoped that simulations could help users learn about and engage with notions of 'competences' in problem solving situations. He was aware of the work of two 'simulation-builders' (Andrea Scharnhorst and Werner Ebeling), and suggested that their specific model approach could be the basis for a simulation. The heuristic potential of models to explain phenomena by assuming a specific set of mechanisms would involve various trade-offs, but it would also hold the advantage of providing a representation with which researchers could explore complex relationships over time. The input to the competence project from Erpenbeck was therefore a body of knowledge about learning, in which learning was specified according to a number of dimensions ('competences').

The core of the simulation-building team was a duo of scholars (Scharnhorst and Ebeling), both trained as physicists, and experienced in the development of various models and simulations. One of them (Scharnhorst) had applied conceptual and formal elements of models to social sciences such as describing technological innovations or the emergence of new ideas. Following their contacts with Erpenbeck, and convinced of the feasibility of such a simulation, they recruited a programmer (Thomas Huesing) who was to program the simulation and design the interface, and an ethnographer (Anne Beaulieu), a colleague of Scharnhorst, who had experience in doing ethnographic work in situations such as laboratories, where learning is an important activity. At this point, the project's aims multiplied. For the programmer, this was an interesting and challenging job that might contribute to his portfolio of expertise. The presence of the ethnographer added a challenge of interdisciplinary work and novelty for all parties. For the physicists on the team, this was an opportunity to explore how their models might be used to convey a social behavior such as learning. At a later point, when the competence project had pretty much run its course, another project around simulations was developing in one of the institutions involved. Matt Ratto, trained as an STS researcher, was the main researchers for this project, and joined in on some of the conversations about the troubled history of the competence project. In trying to recount our many attempts at building a satisfactory simulation to Matt, the various stories we told and the ways in which our accounts contrasted made clear that there were major tensions in the ways we understood the simulations. We recount and illustrate below various efforts to build simulations, and relate them to actors' epistemic commitments (Ratto, 2006).

In the course of discussions between the three authors of this article, a 'back of the envelope' sketch began to take form, as a way of recounting the many efforts in building simulations in the competence project. Sketching enabled us to make sense of the ways in which we disagreed. It also helped us identify the past and present disagreement as variously addressing the meaning, purpose, validity or potential use of simulations. We use our experience to demonstrate the importance of articulating such differences in order to specify and resolve them—or at least, to agree on what we disagree about! The potential of such a chart for developing and putting STS insights to use is explained in the last part of this article. We first consider the kinds of questions that have been raised by researchers around simulation-building, in order to articulate the value of analyzing this area of scientific practice and the potential contributions of science and technology studies to the field of simulations.

**Simulation as Practice**

The characterization of simulation as a particular kind of scientific output has occupied a number of scholars. Interesting discussions have been pursued, as to whether simulations should be considered as scientific experiments, as scientific theories, or as hybrid forms that share features of both concepts, having both a normative and empirical character (Lenhard, Kueppers & Shinn, 2006; Merz & Knuttila, 2006). Many kinds of work can be labeled 'simulations', but generally speaking, a simulation is a partial re-creation of a phenomenon. The phenomena can be recreated through the use of a mathematical model that represents it, but it can also be re-enacted, for example through simulation of behavior in form of games, including war games and role playing. More attention has also been paid recently to practices around simulation and to simulation-building, shifting the view of simulation, from output to process. Ghamari-Tabrizi's work on the development of war-game simulations in the United States demonstrates that running simulations resulted in a special kind of knowledge, grounded in the lived experience of participants rather than in the end-product (Ghamari-Tabrizi, 2000). Similarly, this article documents insights we developed in the course of simulation-building--related to the simulation, but distinct from the 'deliverables' of the project. Focusing on practices also provides insight into the dynamics that shape simulations. Ontological discussions are especially prominent in the building of simulations. (Sundberg, 2006)

In our project there was an appeal to physics, and more specifically, to complexity theory. We will describe these interactions at greater length below, but for now, it is interesting to note that appealing to the physics in our discussions meant asserting the primacy of a mathematical description (an equation). This formulation of a phenomenon was argued to be the most precise and scientific basis for building a simulation. Another study of simulations in nanotechnology (Johnson, 2006), also highlights that if simulations are derived from ‘calculations’, from ‘first principles’, they are considered particularly valuable. Importantly, however, Johnson’s analysis reveals that, when considered in terms of the contextual practice of simulation-building, simulations are best understood as hybrids that are not simply or purely extensions of calculations from models.[1]

While our focus on simulation-building as a practice also emphasizes simulation-building as a chain of heterogeneous processes (Merz and Knuttila, 2006), it would seem that the appeal to an ontological rock bottom is not unique to our experience. If we are to consider that simulations ‘allow actors to hold different conceptions of the same artifact and serve multiple purpose’ and characterize them as ‘unfolding scientific objects’(Merz and Knuttila, 2006: 6), the strength of ontological rhetoric stands in the way of productive discussions about the heterogeneity of simulation-building and about the value of this diversity. This is all the more important because public perceptions of models also distinguish between multiple aspects of models and their functioning (Yearly, 1999). In a setting where public debates increasingly involve models and simulations (Edwards, 1999), understanding and articulating this heterogeneity becomes an important contribution to the politics of knowledge. Debates about the value of various simulations shape simulations as sites where different practices intersect and where different epistemic cultures meet (Galison, 1997; Lenhard, Kueppers & Shinn, 2006). But again, this multiplicity was not always so easy to articulate as a positive quality of simulations. Our process was marked by struggles for purity of form and purpose of simulations, and the diversity of simulations was seen as wasting time. If we are to understand simulation as practice and simulation as a site of diversity and heterogeneity, we must also understand the effect of these appeals to the physics, and the dominance of ontological discussions.

In what follows, we extend the scholarship on these issues by contextualizing these arguments and exploring the implications of ontological discussions in the process of simulation building. Our contribution is therefore germane to the various critiques of simulations that highlight how certain elements may be erased and others privileged in the building of simulation (Helmreich, 2000a). Because we also participated in simulation-building, we are in a privileged position to experience and develop a reflexive stance to simulation building. This is not an easy task. If, as Lansing notes, 'many anthropologists continue to associate any use of mathematics with a simplistic positivism' (Lansing, 2000: 317), we found that many physicists associate verbal description with imprecision, superficiality and an inherently ad hoc relation to 'reality'. Other STS scholars have also noted that it can be difficult to exchange with model-builders about the assumptions on which their simulations or models are built. Lahsen repeatedly signals this issue in her analysis of climate model builders (Lahsen, 2005). She also highlights that while other scholars also have potentially valuable and useful contributions to make to the development of simulation, modelers may not be open to these insights (Lahsen, 2005). Our goal in what follows is therefore is to show how particular emphasis and assumptions in simulation building arise, and how an STS sensibility led us to develop concrete ways of intervening in simulation-building.

## How We Set out to Simulate Competences and Learning

From 2003 to 2005, two of the authors (Scharnhorst and Beaulieu) worked in a project together with a physicist and a computer programmer. The 'competence project', as they grew to call it, was a basic research project meant to inform transformation in the understanding and organization of professional education in Germany. The work was framed by a policy-driven demand for a better understanding of how learning and individual worker development takes place within a knowledge economy. Rather than focus on the role of static qualifications and skills, the project explored a dynamic notion of competence, and developed simulations around this notion. But besides the explicit goal of developing a simulation to represent this notion of competence that was at the forefront of this project, other agendas were also active in shaping interactions during the simulation-building. For Beaulieu and Scharnhorst, this was also an opportunity to explore the interdisciplinary challenge of relating ethnographic observations and simulations. Another element appealed to Scharnhorst and other team members. In the education science framework of the project, competences are conceptualized as dispositions for self-organized learning (Erpenbeck, 1996; Erpenbeck & Rosentiel 2003). For the physicists in the team, this notion of self-organized learning seemed germane to physics' theories of self-organization.[2] Because elements like 'uncertainty' and collective behavior are central to self-organization theory and to discussions about learning and development,[3] to the physicists, this seemed like a promising conceptual connection. It would support the translation of Erpenbeck's notions of learning and competence into simulations based on mathematical models of self-organization with which the physicists were familiar. From this starting point, it seemed to the physicists that the initial task was to develop mathematical models that would incorporate the role of competences in the process of learning, and render the emergence of a self-organizing order.

To transform the framework developed by Erpenbeck and others into a simulation, a number of translations were necessary. In some areas of social science, it is quite common to define behaviors in terms of aspects or features, and this is also the case for Erpenbeck's work. The components of learning (competences and motivations) are described as features, which can be measured according to a questionnaire (also developed by Erpenbeck and colleagues). In contrast to the form of questionnaires and scores, however, the kinds of simulations envisaged were fundamentally spatial. A major translation, therefore, was to transform the scores on each of the relevant features into a spatial representation. In figure 1, we see how scores obtained via a questionnaire can be visualized by means of a spider diagram. The scores for particular features, as ascertained through the questionnaire, are represented as points on the vectors P, A, S and K.

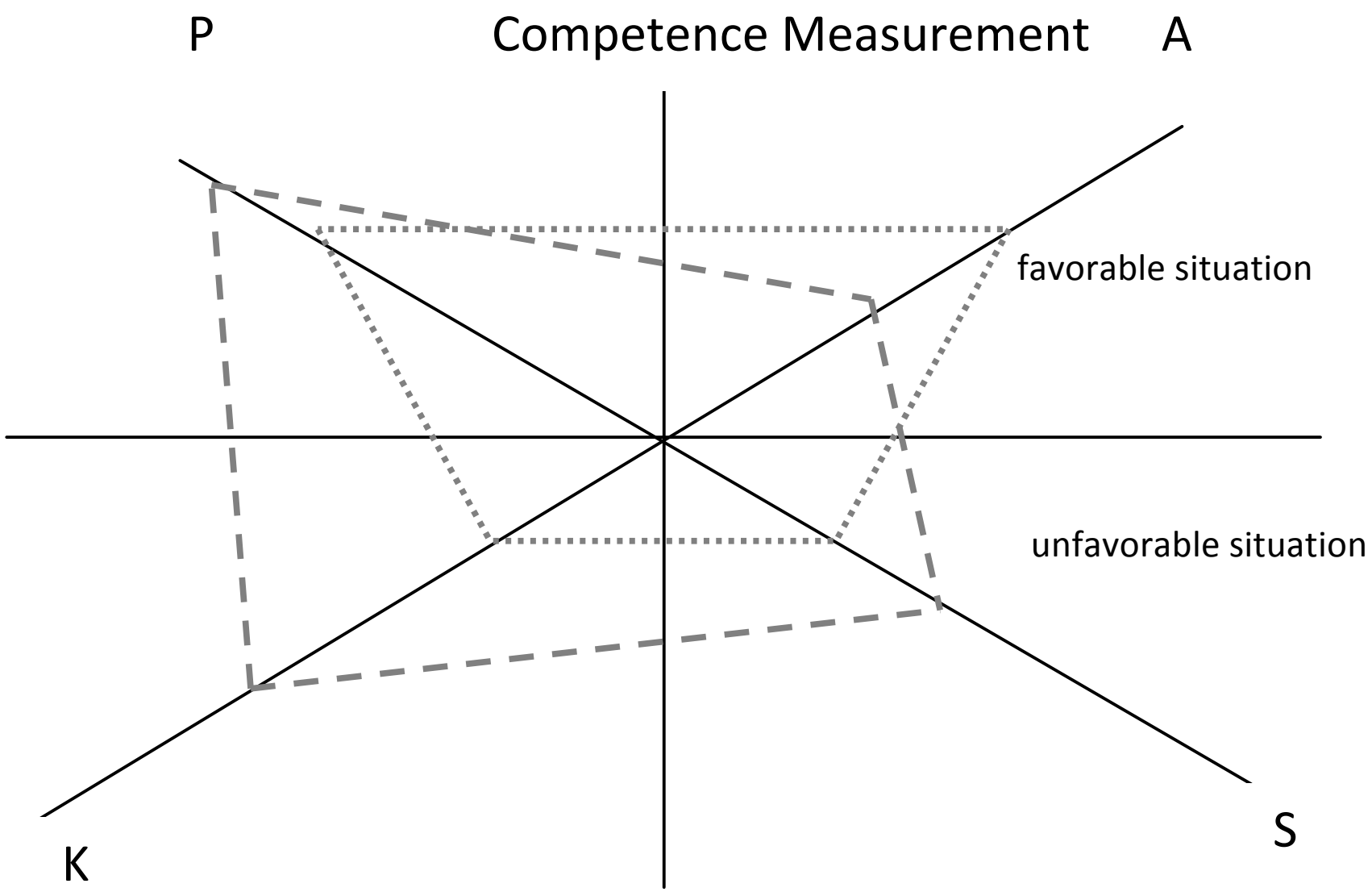

**Figure 1: A space for measuring 4 competences (P, A, S, K). The dashed lines link the scores of an individual in a specific situation. The quantitative scores are thereby rendered spatially. The shape made up by the lines corresponds to particular configurations, which may be more or less favorable to learning.**

A chain is therefore established between concept and spatial representation. In the ‘competence space’ each point represents the actual use of certain competences, and when combined, represent a preferred combination of competences. In the process of learning one can assume that this use pattern changes-- for individuals as well as for the whole group. Trajectories in this space therefore show the development of competences. This representation offers the possibility of building further, more complex representations using a mathematical framework. Via the manipulation of points in a space that have come to stand for competences that support ‘learning’, mathematical modeling can be linked to behavior.

Once the concepts had been given form in a spatial representation, the team’s initial extension of this framework was to transform the static visualization of individual competences are transformed  into a more dynamic visual form that encompassed different configurations (‘scores’) that change over time. Changes in competences could thus be visualized as motion or travel through a landscape where the hills represent “optimal” combinations of competences. When different “searchers” gather together around one peak this means that a certain type of competence spectrum is favored in the group. The competence space represents the development of individual and group skills. The simulation can also be used to show the use of skills and competences to solve certain problem. In this second translation, the competences become mechanisms instead of attributes. The new attributes are features of problems and ideas, and form a problem space. In such an abstract problem space, each location corresponds to a certain state of knowledge reached, or a specific problem solution. Travelling through this space corresponds to the search for new solutions and new ideas (Bruckner et al 1990; Scharnhorst 2001; Weisberg, Muldoon 2009), innovations that could potentially be empirically measured (Scharnhorst 1999).

For the purposes of our discussion here, it is important to note that the formulation ‘agents searching in an abstract landscape can be used to associate the topic to be investigated with a specific genre of simulations.

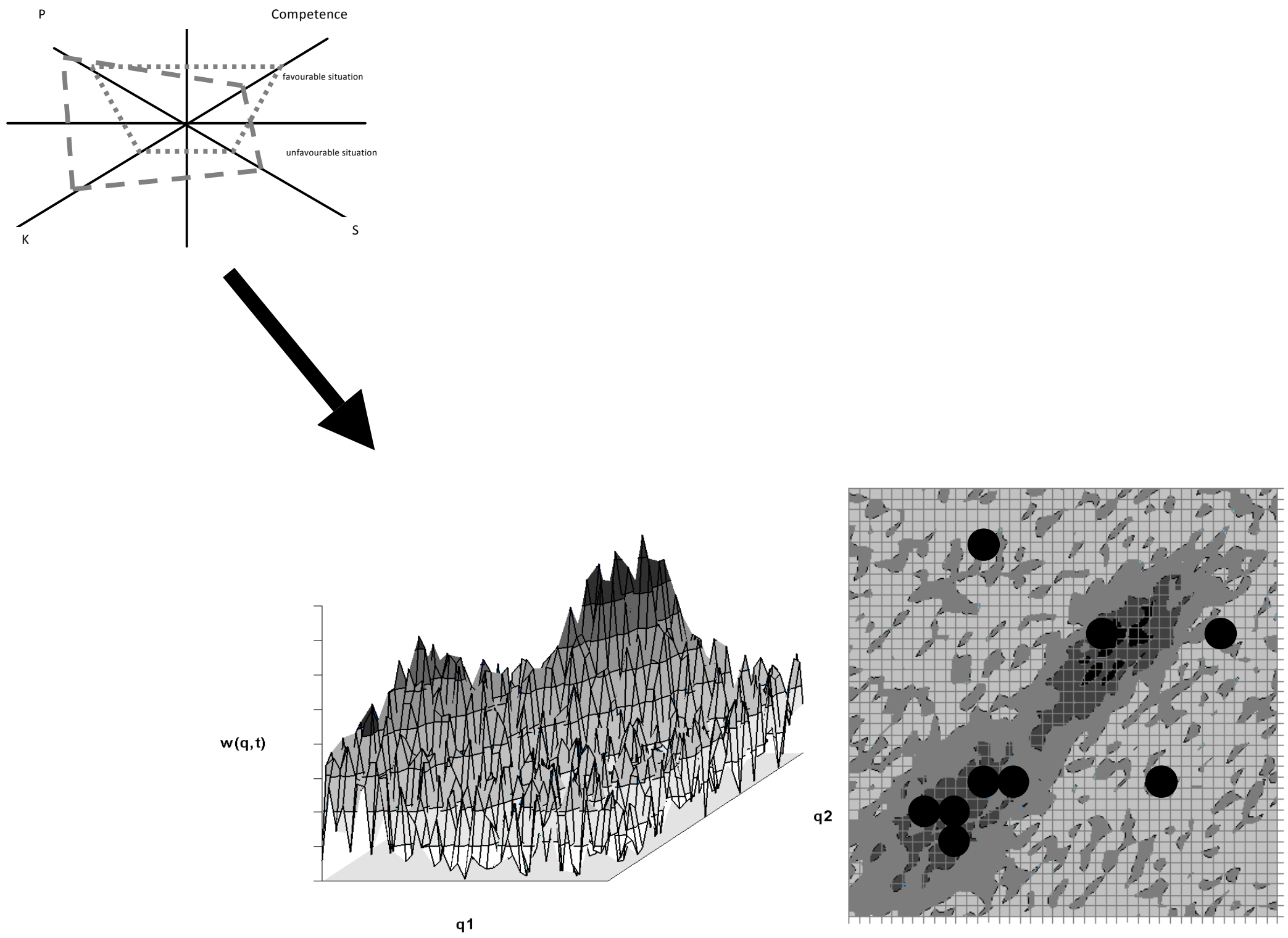

**Figure 2: This figure shows the transformation of one kind of representation into another. The scores (measurement of competences) along 4 dimensions (small figure in upper left corner) are transformed into points on a landscape (shown from side and top view). In the simulation, individuals travel in these landscapes.**

The result is a simulation of individuals experiencing different situations over time, learning or not. The positive consequences of learning are represented as being able to find and climb the hills in the landscape. Referring to our title, this is 'learning in a landscape' in a first sense: we translated and represented learning into a simulation, which takes the shape of traveling through a landscape.

Starting from the phrasing of competence development as 'travel' and moving to the use of competences as instruments for problem solving, the modelers in the team explored the possible mathematical equations and algorithms that could be used to describe the dynamics of movement in a landscape.

*What about ethnography?*

Translations are at the core of building simulations, and many other translations and elaborations had to be specified. But one element that had fallen by the wayside by this point was the idea that ethnographic fieldwork might 'feed' the simulations. Simulation building, as a practice, was in large part a question of anchoring particular aspects of the simulation to features of the behavior to be modeled and to various formal models. It was assumed at the beginning of the project that the role of the ethnographer would be to bring ethnographic observations about competence and professional development into the simulation project. Because there was so little experience on the part of the ethnographer with the practice of simulation building, and equally little familiarity with ethnography on the part of the simulation builders, discussions about possible interactions between ethnographic accounts of competence and formal models proceeded like parallel soliloquies.

What happened next will not be much of a surprise to those familiar with laboratory studies, but came as a rather big and not altogether welcome surprise to the simulation builders: the ethnographer turned her attention to the simulations building team. Questions like 'what is the agency of your agent-based model?' or 'what are you trying to represent with the time dimension?' came up (as in Eglash (1997)), but perhaps most persistent and recurrent were questions about why certain simulations were considered ' good' or successful, and therefore pursued, and why others were abandoned in the very early stages of development. The assumptions that were part of the spatial representation were also critiqued--Why should there be a space? Why should that space be finite? Etc. At times, these questions exacerbated tensions that were already present among members of the project team. What eventually became a fascinating exploration was, for quite a long time, a set of uncomfortable tensions and mounting disagreements. These conversations triggered lasting debates about the different ways of using computer simulations and representing results.

Graphs, videos, interactive game interfaces, visualizations in the form of drawings, and little animated digital images were all eventually used to facilitate communication between members of the project team.

## The Simulations We Built

We provide illustrations of just three of the many simulations that were subsequently built. We chose these because they are most indicative of the work accomplished and of the debates that animated the project. This messiness of the simulation-making process in the project represents, among other things, the complexity of the debate inside the project team. While we present them here as a somewhat linear sequence of events, for the team at that time, the simulations did not emerge progressively. The actual process was in fact, a much more layered, interwoven experience that included loose ends and false starts. As is the case with all representations, this article is a selective, purposive and constitutive account, and one that may become part of varied practices.

The labels given to these selected simulations will be more meaningful to some, but their contrasting appearance should provide readers without much familiarity with particular schools of modeling with a sense of their dissimilarity. A brief overview of the tensions between the different evaluations of the material is provided for each simulation.

### *Swarm simulation*

Swarm models are a favorite tool in complexity theory, are used to model the (coordinated) behavior of a large number of elements (i.e. particles or individuals in populations). We use the term ‘swarm’ to label a numerical algorithm that solves a set of mathematical equations and to refer to its visualization as a swarming movement. Building on work done by Scharnhorst and colleagues, the simulation was developed using a specific framework for modeling self-organizing processes called ‘geometrically-oriented evolutionary theory’. This means that the rules of interaction between agents and landscape are inspired by evolutionary concepts, like adaptation and survival. The movements of agents in the landscape are in turn specified using mathematical equations. The resulting visualization shows agents moving around in a landscape, so that the movement of agents has the appearance of a ‘swarm’, much like that of a swarm of bees. These kinds of simulations are typically used to illustrate emerging group behavior in a complex system. The simulation can be run countless times, and its outcomes can be statistically analyzed. With this simulation, we wanted to see how agents would move from one part of the landscape to the other, from one hill to the next. By changing the parameters that represent the use of competences such as communication or factual knowledge, movement through the landscape could be more or less rapid. Despite the complexity of the underlying mathematical model, this simulation resulted in a somewhat uninspiring visualization. (See figure 3). Most agents finally reached the new maximum in the upper right corner.

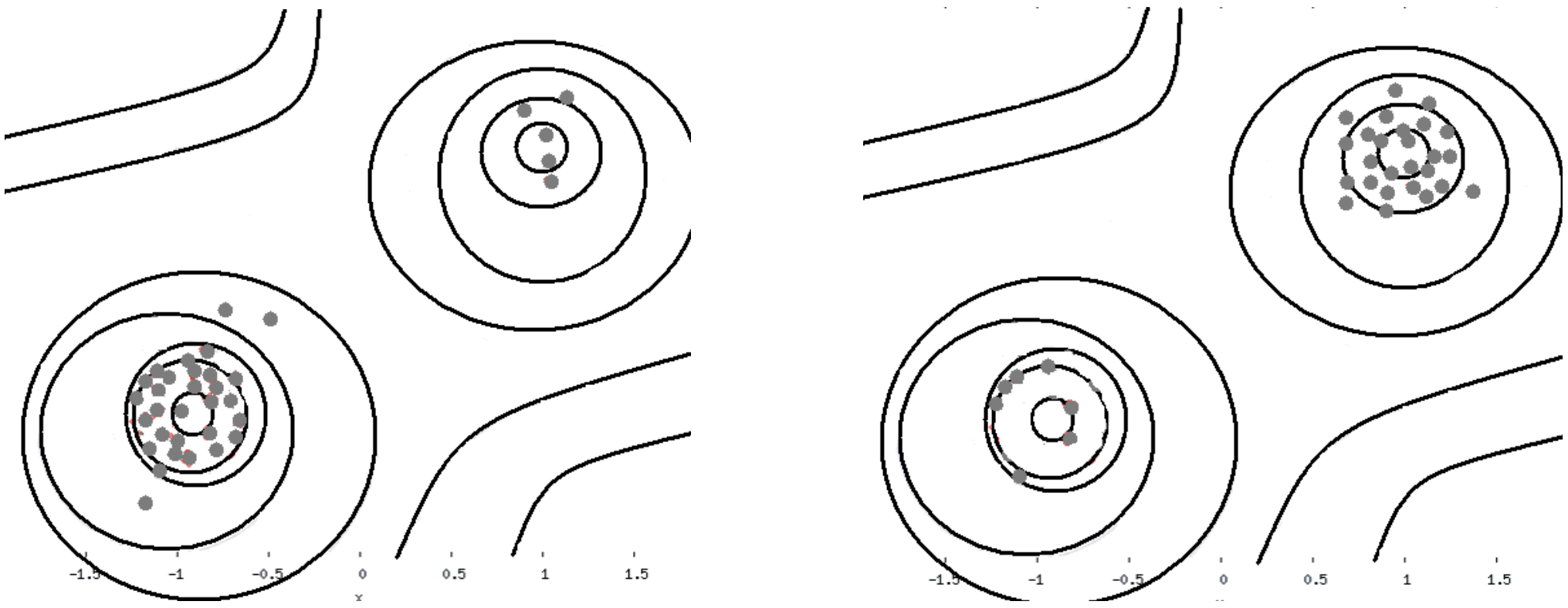

**Figure 3: A top view of the landscape, with hills indicated by contour lines. Agents are shown as small dots. The two screenshots (displayed side by side) show the simulation at two different points in time.**

Some members of the team felt that this simulation successfully incorporated and represented the complexity of the relationships between competences, and between competences and performance, at the level of the underlying model and equations. But little of this sophistication was visible to those unfamiliar with this particular class of mathematical models. While the relationship between the notion of competence and the model underlying the simulation were satisfyingly sophisticated, nearly all other elements of this simulation were disappointing. The result was a visually boring simulation, with points moving in a seemingly trivial manner. If part of the value of using simulations in this project was to engage researchers to think through complex interactions and to support heuristic explorations, this simulation wasn't successful on those counts. Furthermore, specialist programming knowledge, including elements of Fortran and of Linux operating systems, were needed to run the simulation, to change the parameters for agents and to turn the output data file into a video sequence. This further restricted the usability of the simulation.

*Metaphorical simulation*

Following this experience, the team member responsible for computer programming of the simulations took it upon himself to develop a visualization that would better illustrate the main concepts of the competence model. His aim was to create a visualization that would show how competences developed as agents moved between spheres of experience and competence. He also aimed to build a simulation that would be much more interactive than the earlier one, and that would engage users in the exploration of the concepts involved.

This simulation also involved a landscape. The movement of agents, however, was guided by 'rules' rather than by the solving of an equation. The behavior of the agents was rather derived from a verbal description. In this simulation, two large circles represented two different areas of experience. Agents, shown as small circles, first move within the first large circle of experience and eventually travel to the other large circle. The motion of agents is defined according to the impetus each agent derives from its competences.

The result was much more interactive than the swarm simulation. The user could change the number of agents, adjust the level of competences and therefore affect the patterns of travel from one circle to the other. The distance between the circles could also be changed, so as to make travel more or less likely (see figure 4). The support for this simulation was Shockwave Flash, a well-known software suite for web-based animations.

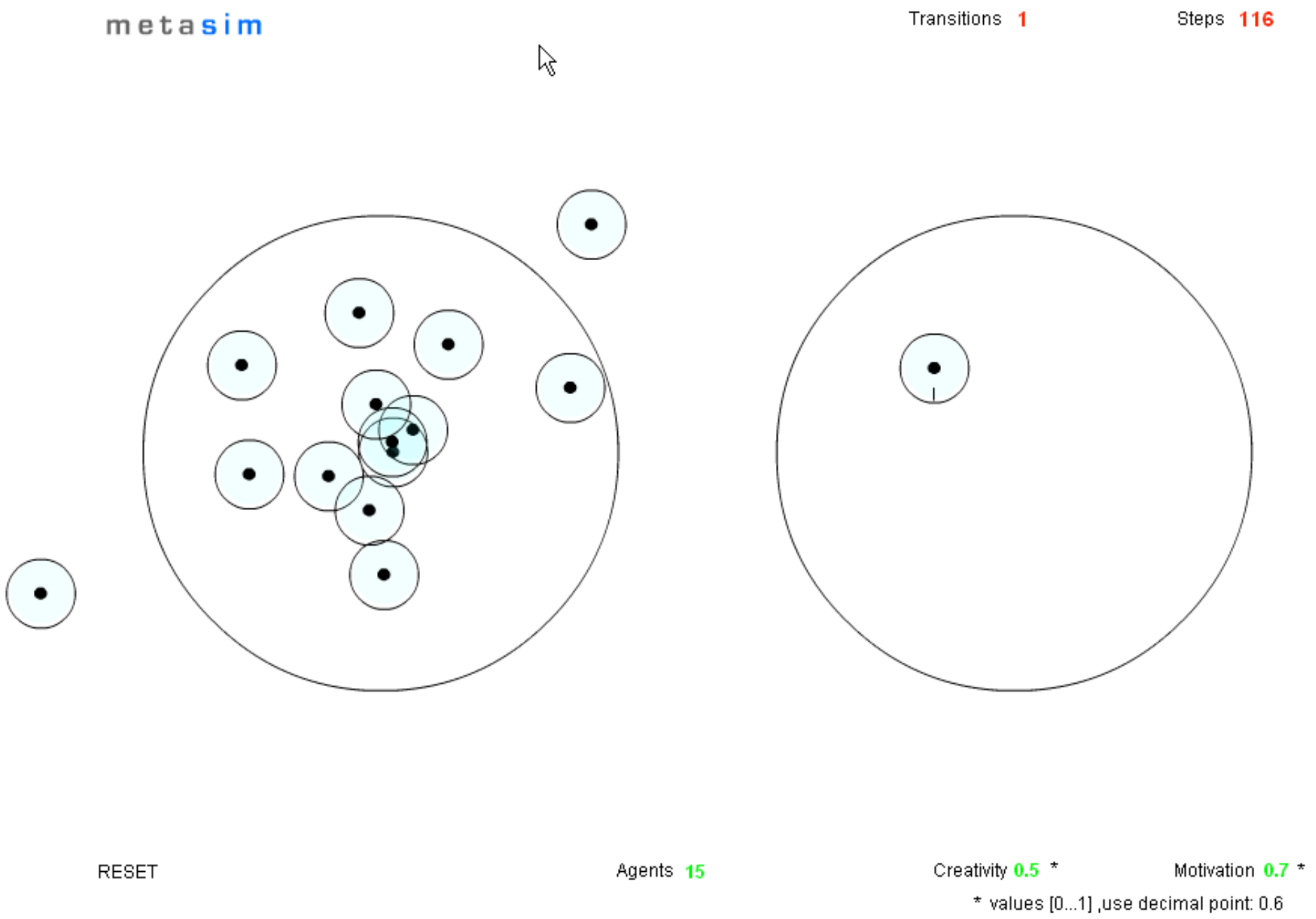

**Figure 4: Snapshot from a run of the interactive 'metaphorical simulation'. Agents are shown as small circles with a nucleus. The two spheres of experience can be seen: the starting sphere is on the left, and agents move from this sphere towards the one on the right.**

Within the team, reactions to this simulation were as strong as they were diverse. Because it did not rely on the specific kind of equation chosen by the physicists, was an initial reaction was to dismiss it out of hand as inadequate. While it did illustrate important features of Erpenbeck's competence theory, its reliance on verbal and conceptual models made it difficult to accept. At best, the physicists argued, we could accept this simulation as a 'game', or as a metaphorical simulation. These terms point to the analogical rather than formally-defined (mathematically-defined) relation between the simulation and underlying model. It was eventually deemed acceptable as an intermediary step towards the possibility of creating an interactive simulation. This was largely due to its form, which opened up the simulation to a non-programming type of interaction with the user, and due to the fact that this simulation made elements of 'competence' more explicit than the swarm simulation.

The metaphoric simulation, in spite —or because— of all the many ways in which it was heretical to the equation-based style of modeling, did open up the discussion to other aspects of simulations. Elements like visual richness, diversity of users, and the exploration of the notion of competence (rather than its being set in stone in terms of its relation to an equation) became important themes of discussion. In the course of these discussions, the team's understanding of simulation also shifted. Rather than seeing simulations as a tool in the service of mathematical theories, the simulations came to be regarded as objects in and of themselves, with significant potential for communicating ideas and concepts. From this point, a greater range of simulation-building strategies were pursued. The metaphoric simulation redirected some of the work towards more interactive forms. Others, feeling that the relation between the set of rules underlying the metaphoric simulation were too arbitrary and vague, returned to complex equations. They did so however, having experienced that more interactive set ups for simulations had advantages.

*Evolino, a Behavioral Simulation*

In this third simulation, the team brought together elements of earlier simulations. The evolutionary model reappeared, with hills and valleys indicating possible situations, while the behavior of agents was made much more explicitly visible and manipulable. This simulation enabled users to play out different scenarios by changing the competences of agents, or by changing the kinds of interactions between group members. In running the simulation, the user could therefore explore how certain relations between individual competences and group processes could alter movement in the landscape. As such, the heuristic possibilities of this simulation were quite high. It seemed to have greater potential for experimenting, rather than solely for demonstrating or visualizing. This version made it possible to play out a situation in which certain types of competence would be highly influential, and to observe the individual and group effects on learning as an outcome.

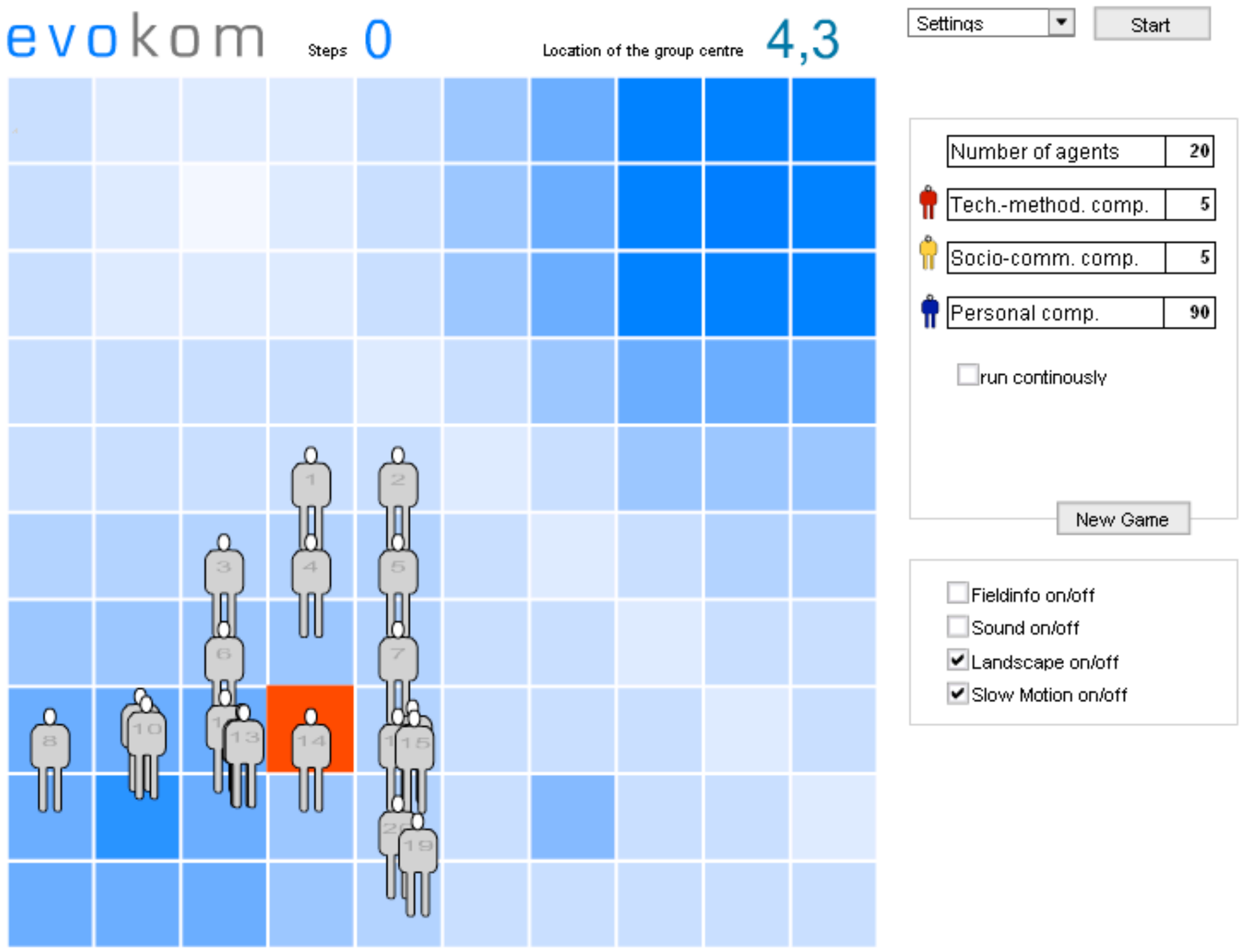

**Figure 5: A screenshot of the Evolino simulation. The variation in the landscape is indicated by shades of blue (grey in this reproduction), with darker areas indicating higher peaks. At the beginning of a run, agents are located at the second lowest of the three peaks, in the bottom left-hand corner. The 'goal', again, is to find the higher hill in the upper right-hand corner. The challenge for this dynamic process is to cross the valley between the hills.**

## Another Strand of the Story: Reflecting on Simulation Building

In the story just told, the research team was tasked with simulating competence and learning, and decided to begin by using mathematical models of evolution, coupled to the metaphor of agents searching across an unknown landscape. The team first created a mathematical model and solved it through the use of stochastic differential equations, outputting the results as a form of swarm dynamics. Deciding this was neither illustrative enough nor sufficiently user-friendly for non-experts, they then used the concepts – but not the mathematics – to code an animated game-like simulation. At a later stage, because of critiques that this last simulation was more of a game than a scientific tool, the team created an interactive simulation that operationalized the concepts through a set of rule-based behavior based explicitly on the evolutionary mathematics, but with a user-friendly graphic cellular automata format.

If we left it there, this would seem to be a straightforward process; the first simulation was explicitly based in proven evolutionary mathematics and therefore carried ontological weight, but was not interactive or 'representationally rich' enough for non-experts; the second simulation, while 'representationally rich', had only a loose or 'metaphoric' connection to the mathematics and therefore was ontologically poor; the combination of these two approaches resulted in the third simulation, one with express links to the mathematics (ontologically rich,) as well as an interactive, and game-like interface (accessible to non-experts). Taken in this way, the story is also reminiscent of Goldilocks' experience with the three bears' porridge –too hot, too cold, just right - and equally a fable.

The actual project was much messier, and, in point of fact, many other simulation prototypes were partially developed and abandoned for various reasons. Several never even reached the eyes of all members of the team. The three simulations detailed above were in fact never considered as 'equal' or part of an overall trajectory – that came with the writing of this article. Each simulation served different and not altogether commensurate purposes inside the research team. The first was used to 'play around' with solutions of mathematical equations in order to discover surprising effects; the second served as a way of convincing members of the team of the need for a more visually rich experience; the third was the means for exploring the creation of simulations that allowed users to discover new perspectives on competence. Moreover, as the simulations developed, the initial perspectives shifted. For example, in the course of the project, the team described the various competence types as both targets of evolutionary processes (in the numerical and metaphoric simulations) and as accompanying mechanisms within learning processes (Evolino). The simulations thus reflected heterogeneity (within the team and over time) about the underlying theories and models, rather than being the result of a purely linear trajectory of trial and error that necessarily resulted, eventually, in the 'best' simulation.

Not surprisingly, there is a close connection between the disciplinary backgrounds and experience of each researcher, and their evaluation of each simulation. The theories of competence and learning, for example, were based on those developed by the philosopher and psychologist (Erpenbeck) who was coordinating the grant framework which funded this work. The physicists (Ebeling and Scharnhorst) had previous experience in modeling evolutionary search processes, using stochastic differential equations, and developing simulations that relied on 'fitness landscapes.' Also trained in social sciences and philosophy of science, one member had experiences in tailoring models towards social theories (Scharnhorst, 2001). The computer programmer (Huesing) had experience developing interactive 'games' and programming user-oriented software. All participants reflected on the relationships, models, and theories promulgated in the project, but the more STS oriented participant (Ratto and Beaulieu) brought a specific critical focus to reflections about disciplinary differences and representational idioms. These differences fuelled the variety of simulations described above, and disagreements remained as to the value of each one of them.

For example, the physicists in the project believed that the swarm dynamics simulation was the richest representation of the underlying self-organizing principles they felt governed competences and learning. This simulation was seen as better because of the way it appeared to make direct use of the formal mathematical model,[4] rather than relying on informal articulations of the concepts or principles that the model was supposed to express. However, the computer programmer and the ethnographer both expressed doubts about the results generated by the simulation and failed to understand them as representing important new insights. They argued that lack of transparency and certain unquestioned assumption seriously limited the simulation, and that mathematical determinism was at play. Equally, since the functioning of the metaphoric simulation involved the programmer's own ad-hoc translation of the competence concepts into computer code, this lack of direct connection to the mathematical model was seen as limiting its applicability. The physicists therefore felt that the behavior of the agents in the metaphorical simulation were neither true to a mathematical model, nor to the evolutionary dynamics that were supposed to be represented by this mathematical model. The validity of the simulation was therefore tightly connected to the use of numerical algorithms.

This connection between validity and numeric algorithms also fostered the physicists' apprehension about what came to be called the 'rule-based' simulation, Evolino. Evolino was seen as a specific numeric translation of the swarm dynamics algorithm into a discrete grid-based game. By taking the different evolutionary functions articulated mathematically in the equations (namely selection, mutation and imitation,) and creating specific corresponding agent rules based upon these functions, members of the team felt the link to evolutionary dynamics would be clearer, thereby increasing the validity of the simulation. Evolino therefore had a more complex articulation of its basic assumptions – from mathematics to rules. It thus claimed most of the validity of the swarm simulation mathematics, with the added value of being an illustrative device for non-physicists.[5] However, to some members of the team, Evolino, like the metaphoric simulation, remained a 'translation' of the more formal swarm simulation-- less valid, but necessary for the further circulation of the original concepts, theories, and ideas.

**From a Back of the Envelope Sketch to a 'Chart'**

This analysis of the experience of the research team highlights the importance of the histories, ways of working and material resources of makers of simulations. These must be taken into account in order to understand where simulations come from and how they are valued. Equally, our experiences have made us very aware of the dangers of too simple a story and the reductive nature of a linear trajectory of development. These are important points to make about the practice of simulation-building, and the insights gained have shaped our approach to simulations in subsequent projects. In conjunction with our reflections on our teams' debates, another type of outcome took shape. A recurring motif in our reflections and negotiations was a 'landscape of simulations', which first appeared as a back of the envelope sketch, and later as a chart (illustrated below). We used this representation in various, increasingly sophisticated forms, in order to articulate tensions and to focus particular debates within the team and between the authors.

As will be clear by now, a key tension characterized the process of simulation-building across the many episodes of the project. On one hand, many different simulations were sketched and elaborated (to different degrees). On the other hand, there was an almost overwhelming tendency to evaluate simulations on an ontological level, in terms of their relationship to a foundational model. As we discussed earlier, if this relationship could be articulated in terms of a mathematical equation, the value of the simulation was considered to be safeguarded, and simulations could proceed from that solid starting point. However, other simulations were differently conceived and articulated, with eventually fruitful consequences for the project. But we found that the golden standard of the 'truth' of simulation was very difficult to denaturalize. We repeatedly faced appeals to 'first principles'--an appeal to the physics that has also been observed by other scholars, as noted above. Our simple sketch started out as a way to make explicit which criteria were being used to evaluate a simulation. Drawing and labeling the axes demanded articulation of the criteria according to which simulations were being evaluated.

The sketch, which developed into something like the figure below (see figure 6), served the purpose of shifting the discussion from 'this simulation is better,' to 'this simulation is better in terms of X or Y.' For example, the version of the chart below shows an x axis that makes explicit a tension between ways of linking models and simulations. The sketch did not magically resolve tensions. But it appeared more and more frequently, and became a way of articulating differences between simulations—differences relative to a specific criterion that was made explicit through consensus about labels.

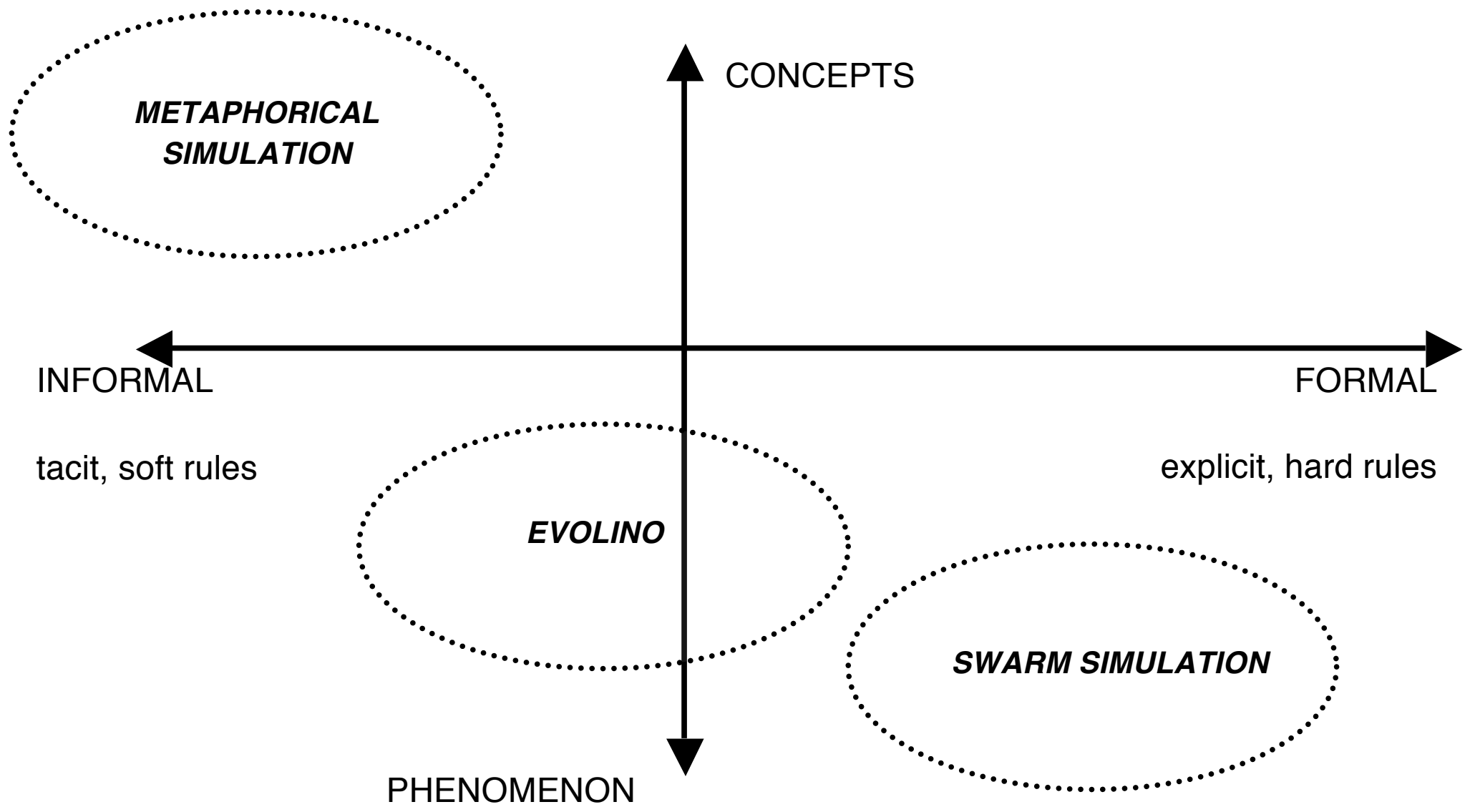

**Figure 6: The various simulations produced in the competence project are placed on the chart, in relation to two axes. The y axis represents the purpose of a simulation, while the x axis represents the methods of description for the model underlying the simulation.**

In this project, the dimension of formal/informal appeared especially frequently, as the way to articulate (and open up to debate) the ontological hierarchy that was at work in assigning greater value to simulations based on mathematical models (equations) than to simulations based on concepts (verbal descriptions). This particular label is the object of a fragile consensus, and still the object of discussion in the writing of the final version of this text.

But it is neither the normative nor finalized status of this chart that is significant. Rather, it is the way the chart enables the articulation of the criteria for evaluation. This is learning in a landscape in a second sense: in developing this representation of simulations in a landscape, we learned to better debate simulations. It was at times a great struggle to establish that certain simulations could be better according to certain criteria, or be better-suited for certain purposes. The need to make assumptions explicit is not always equally felt by all team members, and it is in such situations that a particular mode of expression, such as this chart, can be of use. This way of representing simulations made explicit that there were different ways of valuing them. This is not a trivial achievement since it affects how the direction of a project is determined.

In other discussions, however, different versions of the chart helped to interrogate and make explicit other aspects of simulation, such as issues of closure and open-endedness. This particular contrast is especially important in the encounter of modeling and ethnographic modes of research (Beaulieu, Scharnhorst and Wouters, 2007).

In figure 7, we show a possible variations of our initial sketch, and suggest potential dimensions which might enter discussions of simulation in the course of building them. In presenting this sketch-turned-into-a-chart, we feel the tension between reifying how simulations should be understood, and offering our experience as a potentially powerful contribution to developing a more reflexive approach to simulation building.

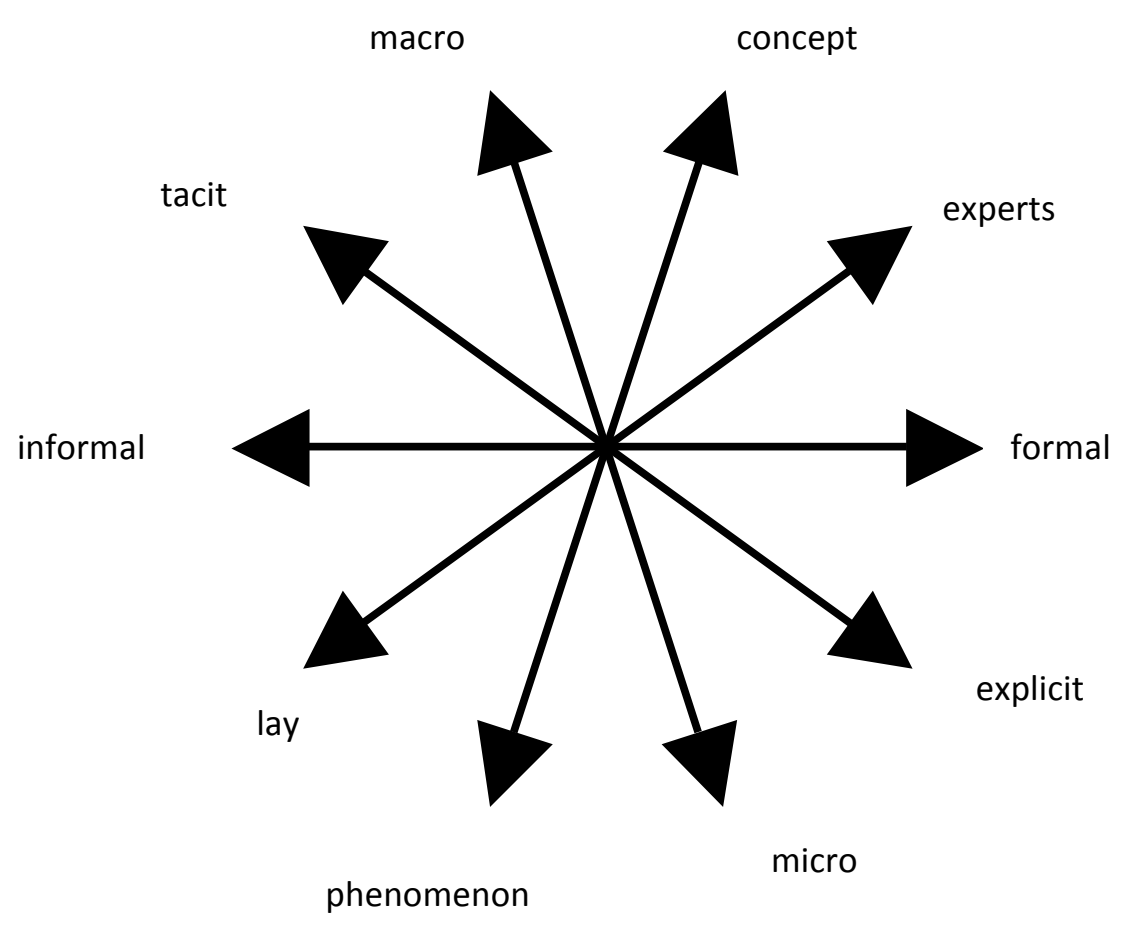

**Figure 7: Possible dimensions of simulations. Our chart focuses on two specific dimensions that were especially important in shaping debates in our own simulation -building project. Several others have been identified in this paper, and many others can be imagined.**

Our chart can be seen as an instance of pictures that simply 'simulate a passage from a literary surface to a witnessable object or practical work space' (Lynch, 1991: 18). We are aware that we undertake to spatialize our arguments—much in the same way that learning became something to be represented in a landscape. To deploy STS insights about the value of specifying criteria in the evaluation of knowledge, we transformed them into the idiom of our colleague simulation-builders. Representational features, such as labels, vectors, structural axes and boundaries, all contribute to creating different impressions, even in sociological figures (Lynch, 1991). The landscape enabled us to articulate not only differences between simulations, but also their potential co-existence. If nothing else, our use of a populated landscape to elaborate our arguments was successful in increasing the impression of relevance for our simulation-building audiences. Our appeal to a visual rhetoric is motivated by pragmatic rather than scientistic aspirations, and we want to stress that the chart has worked best when used as a prop, as a starting point to be modified in the course of talking and sketching rather than as reified representation.

## Simulation as Reduction, Representation and Intervention

Simulations always involve selection and abstraction, no matter the composition of the team or the domain to which they are applied. As a result of the analysis of simulation-building presented here, however, we are in a position to make two important points in relation to concerns that the use of simulations carries the risk of embracing reductionism or positivism.

First of all, we wish to point out that there may also be a risk involved in NOT doing this. Within anthropology, STS or other interpretative fields, 'cultural and historical processes', as well as human agency, are no less expressed in 'idioms' than when represented in simulations. By challenging our representations of 'cultural and historical processes', the competence project brought to light our investments in particular representations. For example, working on these simulations involved a number of clashes between expectations of 'delivering data' and 'pursuing fieldwork'. While these at first seemed like clashes between positivistic approaches and interpretative ones, prolonged discussions of these expectations and of ways of working clarified that the issue was rather both more specific and more fundamental than it seemed at first. In trying to incorporate ethnographic material into a simulation, we came to focus on and articulate the differences in the kinds of 'closure' that are achieved in physics and in ethnographic fieldwork, and on the radically different timeframes in which this occurs. Modeling could not begin without a description of a phenomenon, but, equally, fieldwork could not proceed without an open-ended view of what a phenomenon might be.[6] By being involved in simulation-building, assumptions about fieldwork became explicit, because they were challenged by requests for particular kinds of representations of fieldwork. Friction between idioms reveals assumptions that remain invisible when sticking to overly familiar modes of representation.

As we see it, the risk for researchers working in interpretative traditions to engage with simulation-building, is *also* the risk of gaining a better understanding of what is gained and what is lost, of what we can possibly compromise of our usual ways of describing and representing, and what is non-negotiable. Given that simulations can play an important role in the representation of complex multi-causal phenomena, a class of systems of prime importance for the humanities (Hayles, 2005), we see this as a risk well worth taking.

Second, we also hope to have shown that simulation building is far from being a hegemonic 'technical representation', but rather a modulated set of representations which, in their diversity, allow for a moderately diverse range of encounters. Even in our small project team, many simulations were generated, each putting forth very different elements and highlighting different ways of knowing. Of course, part of the story we told also highlights that this range is limited in certain ways, and that there is a strong tendency to consider simulations according to particular hierarchies—in our case, ones that fit with an epistemic culture of physics. But this can only change as researchers from the social sciences engage in, develop, and influence simulation-building practices.

Our chart is proposed as a contribution to further facilitate and enhance such encounters. We suggest that this kind of representation can be a useful and potentially new kind of contribution of STS to scientific practice.

Besides these developments, an enduring contribution of STS has been to make visible various kinds of work that are usually excluded from accounts of science. Our account contributes to this larger project by making visible the tensions and activities involved in combining roles as participant and analyst. Our role as both makers of simulations and as observers of our own project results in valuable and arguably unique insights, but it also requires close attention. For example, while our immersion in the processes of doing simulation gave us unparalleled access to the ongoing 'messy' practices involved in this work, this immersion also generated particular attachments to our own sense of how the project progressed, and made it particularly difficult to 'report' on this work. We have attempted to maintain 'epistemological reflexivity,' one of the important contributions of ethnographic approaches to studies of scientific and technological processes (Forsythe, 1997), a practice that has taken the form of signalling compromises and tensions (also in the writing of this piece).

The dual roles of analyst and participant are also increasingly played by STS scholars working with a mandate for intervention, in the course of constructive technology assessment exercises, or as part of large endeavors around genomics, nanotechnology or cyberinfrastructure. Our hope is therefore that this article may contribute to document simulation-building as a practice, that the chart may be a concrete contribution to stimulate and focus discussions about ways of knowing, and that we may have demonstrated how STS insights can be deployed in the course of combining dual roles of participant and analyst.

**Notes**

The simulation-building work described in this manuscript was funded by the Federal Ministry for Education and Research (Germany) and the European Social Fund and while the work of Matt Ratto was funded by the Dutch Organisation for Scientific Research (NWO) through the project Dissimilar Simulation: the Epistemics of Simulation in the Humanities, in the framework program 'Culturele vernieuwing en de grondslagen van de geesteswetenschappen'. We are grateful to the other members of the simulation project, to our colleagues for discussions on these matters, and to John Erpenbeck who encouraged us to work on the interactive aspect of the simulations. We would also like to thank the editors and reviewers, Marcel Boumans, Guenter Kueppers, Sabina Leonelli, Kyriaki Papageorgiou, as well as members of the Virtual Knowledge Studio for their detailed comments on earlier versions of this article. The simulations developed in this project can be found at http://evolino.virtualknowledgestudio.nl/english/files.html .

1. In assessing the usefulness of simulations for social science, it is precisely the possibility that simulations can be 'more' than mathematical formulations that makes them appealing (Moretti, 2002).

2. In particular, self-organization theories developed in physics frame this work (Nicolis & Prigogine, 1977, 1989; Feistel & Ebeling 1989).

3. Uncertainty is important for two reasons. First, it makes explicit that learning often occurs in situations where the goal is shaped during the process (Erpenbeck & Heyse, 1999). Second, uncertainty is an implicit part of the 'real world' context in which competence is exercised, where, for example, the problem of unemployment and the need for flexible work forces remain major policy issues.

4. However, it is important to make explicit that even in the case of the swarm simulation, some 'massaging' of the equations was required in order to make them work within the swarm software framework. This type of 'articulation work' (Strauss, 1988; Fujimura, 1987), often referred to as 'tuning', has been documented in other examples of mathematical modelling and simulation. (E.g. Kueppers & Lenhard, 2005; Winsberg, 2006).

5. Actually, Evolino has turned out not to be particularly suitable for the kinds of model-testing (e.g. the statistical analysis of high numbers of individual simulation 'runs',) that physicists and other simulation users often require for legitimacy. The computer language (shock wave flash) requires quite a lot of memory and as a web program does not allow the storage of results. The simulation therefore tends to crash if it runs for too long or too often.

6. This and other insights are detailed in another article (Beaulieu, Scharnhorst & Wouters, 2007).

**References**

Bal, Roland, Wiebe Bijker & Ruud Hendriks (2004) 'Democratization of Scientific Advice', *British Medical Journal* 329(7478): 1339-41.

Beaulieu, Anne & Paul Wouters (in the press) 'E-research as Intervention', in Nick Jankowski (ed.), *e-Research: Transformations in Scholarly Practice* (London: Routledge).

Beaulieu, Anne, Andrea Scharnhorst & Paul Wouters (2007) 'Case Study: a Middle-range Interrogation of Ethnographic Case Studies in the Exploration of E-science', *Science, Technology & Human Values* 32(6): 672-92.

Edwards, Paul N. (1999) 'Global Climate Science, Uncertainty and politics: Data-Laden Models, Model Filtered Data', *Science as Culture* 8(4): 437-72.

Eglash, Ron (1999) *African Fractals: Modern Computing and Indigenous Design* (New Brunswick, NJ: Rutgers University Press).

Erpenbeck, John (1996) 'Synergetik, Wille, Wert und Kompetenz', *Ethik und Sozialwissenschaften* 7(4): 611-13.

Erpenbeck, John & Volker Heyse (1999) 'Kompetenzbiographie - Kompetenzmillieu - Kompetenztransfer: Zum biologischen Kompetenzerwerb von Fuehrungskraeften der mittleren Ebene, nachgeordneten Mitarbeitern und Betriebsräten', in John Erpenbeck and Voler Heyse (eds), *QUEM-report Schriften zur beruflichen Weiterbildung* (Berlin: QUEM) 62: 106-40.

Erpenbeck, John & Lutz von Rosentiel (2003) 'Einfuehrung', in John Erpenbeck and Lutz von Rosenstiel (eds), *Handbuch Kompetenzmessung* (Stuttgart: Schäfer-Poeschel): IX-XL.

Feistel, Rainer & Werner Ebeling (1989) *Evolution of Complex Systems* (Dordrecht: Kluwer).

Forsythe, Diana (1997) 'Representing the User in Software Design', unpublished manuscript, <http://www.stanford.edu/dept/HPS/forsythe.paper.html>.

Fujimura, Joan (1987) 'Constructing "Do-able" Problems in Cancer Research: Articulating Alignment', *Social Studies of Science* 17(2): 257-93.

Galison, Peter (1997) *Image and Logic: A Material Culture of Microphysics* (Chicago: University of Chicago Press).

Ghamari-Tabrizi, Sharon (2000) 'Simulating the Unthinkable: Gaming Future War in the 1950s and 1960s', *Social Studies of Science* 30(2): 163-223.

Hayles, N. Katerine (2005) *My Mother was a Computer: Digital Subjects and Literary Texts* (Chicago: University of Chicago Press).

Helmreich, Stefan (2000a) 'Digitizing 'Development' ', *Critique of Anthropology* 20(3): 249-65.

Helmreich, Stefan (2000b) 'Power/Networks', *Critique of Anthropology* 20(3): 319-20.

Johnson, Ericka (2007) 'Surgical Simulators and Simulated Surgeons: Reconstituting Medical Practice and Practitioners in Simulations', *Social Studies of Science* 37(4): 585-608.

Johnson, Ann (2006) 'Institutions for Simulations: The Case of Computational Nanotechnology', *Science Studies* 19(1): 35–51.

Kueppers, Guenter, & Johannes Lenhard (2005) 'Validation of Simulation: Patterns in the Social and Natural Sciences', *Journal of Artificial Societies and Social Simulation* 8(4): 3 <http://jasss.soc.surrey.ac.uk/8/4/8.html>.

Lahsen, Myanna (2005) 'Seductive Simulations: Uncertainty Distribution around Climate Models', *Social Studies of Science* 35(6): 895-922.

Lansing, Steve (2000) 'Foucault and the Water Temples: a reply to Helmreich', *Critique of Anthropology* 20(3): 309-18.

Lenhard, Johannes, Guenter Kueppers & Terry Shinn (eds) (2006) *Simulation: Pragmatic Construction of Reality* (Dordrecht: Kluwer Academic Publishers).

Lynch, Michael (1991) ‘Pictures of Nothing? Visual Construals in Social Theory’, *Sociological Theory* 9(1): 1-21.

Marres, Noortje & Richard Rogers (2000) ‘Landscaping Climate Change: A mapping technique for understanding science & technology debates on the World Wide Web’, *Public Understanding of Science* 9(2): 141-63.

Matthies, Michael, Horst Malchow & Juergen Kriz (eds) (2001) *Integrative Systems Approaches to Natural and Social Dynamics* (Heidelberg: Springer).

Merz, Martina (2006) ‘Embedding Digital Infrastructure in Epistemic Culture’, in Christine Hine (ed.), *New Infrastructures for Knowledge Production: Understanding E-Science* (Hershey: Idea Group Inc.): 99-119.

Merz, Martina & Talja Knuttila (2006) 'Computer Models and Simulations in Scientific Practice', *Science Studies* 19(1): 3-11.

Mogoutov, Andrei, Alberto Cambrosio & Peter Keating (2005) ‘Making Collaboration Networks Visible’, in Bruno Latour and Peter Weibel (eds), *Making Things Public. Atmospheres of Democracy* (Cambridge, MA: MIT Press): 342-45.

Moretti, Sabrina (2002) 'Computer Simulation in Sociology: What Contributions', *Social Science Computer Review* 20(1): 43-57.

Nicolis, Gregoire & Ilya Prigogine (1977) *Self-organization in non-equilibrium systems* (New York: Wiley).

Nicolis, Gregoire & Ilya Prigogine (1989) *Exploring complexity* (New York: WH Freeman).

Ratto M (2006) ‘Epistemic Commitments and Archaeological Representation’ in Luiz Oosterbeek & Jorge Raposo (eds), *XV Congrès de l’Union Internationale des Sciences Préhistoriques et Protohistoriques. Livre des Résumés* 1: 60 <http://www.uispp.ipt.pt/UISPPprogfin/Livro2.pdf>.

Scharnhorst, Andrea (2001) 'Constructing Knowledge Landscapes within the Framework of Geometrically Oriented Evolutionary Theories', in Michael Matthies, Horst Malchow & Jorgen Kriz (eds), *Integrative Systems Approaches to Natural and Social Sciences Systems Science 2000* (Berlin: Springer): 505-15.

Sperschneider, Werner & Kirsten Bagger (2003) ‘Ethnographic Fieldwork under Industrial Constraints: Toward Design-in-Context’, *International Journal of Human Computer Interaction* 15(1): 41-50.

Strauss, Anselm (1988) ‘The Articulation of Project Work: An Organizational Process’, *The Sociological Quarterly* 29(2): 163-78.

Suchman, Lucy (1994) ‘Working Relations of Technology Production and Use’, *Computer Supported Collaborative Work (CSCW)* 2(1-2): 21-39.

Suchman, Lucy (2002) ‘Located Accountabilities in Technology Production’, *Scandinavian Journal of Information Systems* 14(2): 91-105.

Sundberg, Mikaela (2006) 'Credulous Modellers and Suspicious Experimentalists? Comparison of Model Output and Data in Meteorological Simulation Modeling', *Science Studies* 19(1): 52-68.

Yearley, Stephen (1999) 'Computer Models and Public's Understanding of Science: a Case Study Analysis', *Social Studies of Science* 29(6): 845-66.

Michael Weisberg & Ryan Muldoon (2009) ‘Epistemic Landscapes and the Division of Cognitive Labor’. *Philosophy of Science* 76 (2): 225-252.

Winsberg, Eric (2006) ‘Models of Success vs. The Success of Models: Reliability without Truth’, *Synthese* 152(1): 1–19.

Zuiderent-Jerak, Teun & Casper Bruun-Jensen (2007) ‘Unpacking "Intervention" in Science and Technology Studies’*, Science as Culture* 16(3): 227-35.

Anne Beaulieu is Senior Research Fellow and Deputy Programme Leader of the Virtual Knowledge Studio in Amsterdam. Her work focuses on two issues: the role of databases and networks in knowledge creation, and the development of new approaches to study cultural and social processes in mediated settings (virtual ethnography). She is currently working on a project entitled Network Realism, an ethnographic study of the creation of knowledge via databases of images on the web.

Address: The Virtual Knowledge Studio for the Humanities and Social Sciences - VKS
Royal Netherlands Academy of Arts and Sciences, Cruquiusweg 31, 1019 AT Amsterdam
The Netherlands; fax: +31 20 850 0277, email: anne.beaulieu@vks.knaw.nl

Matt Ratto is assistant professor in the Faculty of Information, University of Toronto. His work brings a science studies perspective to critical studies of digital media. Recent publications include ‘*Ethics of Seamless Infrastructures: Resources and Future Directions*’ and ‘*A Practice-Based Model of Access for Science: Linux Kernel Development and Shared Digital Resources*.’ He is currently working on a book manuscript entitled *Critical Making* that examines how designers, scholars, and technologists use material practices of "making" to conceptualize and express knowledge about the world.

Address: Faculty of Information Studies, University of Toronto, 140 St. George Street, Toronto, Ontario M5S 3G6; fax: +1 416 978-5762; email: matt.ratto@utoronto.ca

Andrea Scharnhorst is Senior Research Fellow the Virtual Knowledge Studio in Amsterdam, where she leads a collaboratory on modeling and simulation in the humanities and social sciences. Her work focuses on the use of mathematical models (in particular models of self-organization, evolution and complex systems) as heuristic tools. She is concerned with issues of boundary conditions and dynamic processes behind systemic innovations, and recently edited a book (together with Andreas Pyka), *Innovation Networks – New Approaches in Modelling and Analyzing*. She is engaged in translating and bridge-building between different knowledge areas, in particular between the natural sciences and social sciences and humanities.

Address: The Virtual Knowledge Studio for the Humanities and Social Sciences - VKS
Royal Netherlands Academy of Arts and Sciences, Cruquiusweg 31, 1019 AT Amsterdam
The Netherlands; fax: +31 20 850 0277; email: andrea.scharnhorst@vks.knaw.nl